\documentclass[preprint,12pt,3p]{elsarticle}

\usepackage{graphicx}                           
\usepackage{pgfplots}
\pgfplotsset{compat=1.18}
\usepgfplotslibrary{groupplots}
\usepgfplotslibrary{colormaps}
\usepgfplotslibrary{colorbrewer}
\usetikzlibrary{math}
\usepgfplotslibrary{external}
\usepackage[exponent-product=\cdot]{siunitx}    
\usepackage{amsmath,amssymb, mathtools}         
\usepackage{scrlayer-scrpage}                   
\usepackage[english]{babel}                     
\usepackage{booktabs}                           
\usepackage{caption, subcaption}
\usepackage{bm}
\usepackage{url}
\usepackage{xcolor}
\definecolor{col1}{rgb}{0.1, 0.2, 0.7}
\definecolor{col2}{rgb}{0.8, 0.1, 0.2}
\definecolor{col3}{rgb}{0.3, 0.7, 0.3}
\definecolor{col4}{rgb}{0.6, 0.2, 0.6}
\definecolor{col5}{rgb}{1.0, 0.6, 0.1}
\usepackage[hidelinks]{hyperref}

\newcommand{\vc}[1]{\bm{#1}}
\newcommand{\mt}[1]{\bm{#1}}
\renewcommand*{\div}[1][]{\nabla\cdot #1}
\newcommand{\grad}[1]{\nabla #1}
\newcommand{\pd}[2]{\frac{\partial #1}{\partial #2}}
\renewcommand*{\vec}[1]{\bm{#1}}

\journal{Computer Physics Communications}

\begin{document}

\begin{frontmatter}

\title{A P-Adaptive Hybridizable Discontinuous Galerkin Spectral Element Method for Electrostatic Particle-in-Cell Simulations}
\author[IRS,1]{Tobias Ott}
\ead{ottt@irs.uni-stuttgart.de}
\author[BOLTZ]{Stephen Copplestone}
\author[IRS]{Marcel Pfeiffer}

\affiliation[IRS]{organization={Institute of Space Systems, University of Stuttgart},
            addressline={Pfaffenwaldring 29, 70569 Stuttgart},
            country={Germany}}
\affiliation[BOLTZ]{organization={boltzplatz - numerical plasma dynamics GmbH},
            addressline={Schelmenwasenstraße 32, 70567 Stuttgart},
            country={Germany}}
\affiliation[1]{Corresponding author}

\begin{abstract}
This paper presents a p-adaptive high-order hybridizable discontinuous Galerkin spectral element method (HDG-SEM) for solving the Poisson equation in electrostatic plasma simulations using particle-in-cell (PIC) schemes.
This approach enables element-local refinement of the polynomial degree, concentrating computational effort specifically in regions with strong gradients.
Thus, the method significantly reduces the global number of degrees of freedom compared to uniform high-order methods.
The proposed method is implemented in the open-source framework \textsc{PICLas} and validated through a series of benchmark test cases, including a dielectric sphere and a one-dimensional plasma sheath.
Finally, a two-dimensional axisymmetric simulation of an ion optic demonstrates the method's capability to efficiently model complex plasma phenomena but also highlights current limitations.
\end{abstract}

\end{frontmatter}

\section{Introduction}
The simulation of rarefied plasmas is essential for numerous modern engineering applications, ranging from electric space propulsion systems to vacuum arc discharges and semiconductor manufacturing.
In such highly rarefied, non-equilibrium conditions, fluid approximations often break down, and the Vlasov equation must be used to accurately describe the kinetic behavior of the plasma.
A central computational challenge in these simulations is the accurate and efficient calculation of the mean electric field, which, under the electrostatic approximation, requires solving the Poisson equation at each time step.
\\
The Particle-in-cell (PIC) method is the most widely used approach for solving the Vlasov-Poisson system numerically.
In PIC schemes, macro-particles are used to represent the probability density function of each species, while the electromagnetic or electrostatic fields are calculated on an underlying Eulerian computational grid.
The simulation cycle involves interpolation or projection of the particle charges onto the grid, solving the field equations, and then interpolating the fields back to the particle positions to advance their trajectories.
\\
Traditionally, PIC codes have relied on low-order finite difference methods for the field solver \cite{birdsallPlasmaPhysicsComputer1985,hockneyComputerSimulationUsing2021}.
However, phenomena such as plasma sheaths or strongly localized charge distributions introduce sharp spatial gradients in the electric potential.
Accurately resolving these features with low-order methods requires a very fine mesh, which leads to high computational costs and memory demands.
High-order numerical methods, such as the hybridizable discontinuous Galerkin (HDG) method \cite{cockburnSuperconvergentLDGhybridizableGalerkin2008,cockburnHybridizableDiscontinuousGalerkin2009,cockburnDerivationHybridizableDiscontinuous2009,cockburnUnifiedHybridizationDiscontinuous2009}, offer a promising alternative for solving the Poisson equation in PIC simulations \cite{jacobsHighorderNodalDiscontinuous2006,pfeifferParticleCellSolverBased2019}.
\\
In this work, we present a p-adaptive hybridizable discontinuous Galerkin spectral element method (HDG-SEM) for the solution of the Poisson equation in electrostatic PIC simulations.
By enabling element-local refinement of the polynomial degree, the solver focuses computational resources strictly on regions with high spatial gradients, significantly reducing the overall number of degrees of freedom while maintaining high accuracy.
\\
The proposed P-adaptive HDG-SEM is implemented within the open-source simulation framework \textsc{PICLas} \cite{munzCoupledParticleInCellDirect2014,fasoulasCombiningParticleincellDirect2019}, a massively parallel code for coupled PIC and direct simulation Monte Carlo (DSMC) simulations of rarified plasma and gas flows.
\\
The remainder of this paper is organized as follows:
Section 2 outlines the physical models described by the Vlasov-Poisson system.
Sections 3 and 4 detail the numerical discretization and implementation, emphasizing the formulation and data structures required for the p-adaptive HDG-SEM.
In Section 5, the method is validated through benchmark test cases, including a dielectric sphere, a one-dimensional plasma sheath, and a two-dimensional axisymmetric simulation of an ion optic, demonstrating its accuracy and computational efficiency.

\section{Theory}
The kinetic evolution of a species $\alpha$ in a collisionless, unmagnetized plasma is governed by the Vlasov equation:
\begin{equation}
    \pd{f_\alpha}{t} + \vc{v}\cdot\grad{f_\alpha} + \frac{q_\alpha}{m_\alpha}\left(\vc{E} + \vc{v}\times\vc{B}\right)\cdot\pd{f_\alpha}{\vc{v}} = 0,\qquad (\vc{x},\vc{v})\in\Omega\times\mathbb{R}^3,
\end{equation}
where $f_\alpha(\vc{x},\vc{v},t)$ is the probability density function in phase space, and $q_\alpha$ and $m_\alpha$ denote the species charge and mass, respectively.
The acceleration on physical particles is caused by the Lorenz force, which depends on the electric field $\vc{E}$.
In this work, we consider the electrostatic approximation, assuming that induced magnetic fields are negligible.
Consequently, the electric field can be expressed as the negative gradient of a scalar electric potential $\varphi$, such that $\vc{E} = -\grad{\varphi}$.

In the presence of dielectric materials, the electric field is related to the electric displacement field $\vc{D}$ by $\vc{D} = \varepsilon\vc{E}$.
The electric potential is obtained by solving the Poisson equation, which can be written as a first-order system:
\begin{alignat}{2}
    \frac{1}{\varepsilon}\vc{D} +\grad{\varphi} &= 0, \qquad &\vc{x}\in\Omega,\label{eq:poisson1}\\
    \div{\vc{D}} &= \rho, \qquad &\vc{x}\in\Omega,\label{eq:poisson2}
\end{alignat}
where $\varepsilon=\varepsilon_0\varepsilon_r$ is the absolute permittivity, which is the product of the vacuum permittivity $\varepsilon_0$ and the relative permittivity $\varepsilon_r$, and $\rho$ is the charge density.
The system is closed with Dirichlet and Neumann boundary conditions:
\begin{alignat}{2}
    \varphi &= \varphi^D, \qquad &\vc{x}\in\partial\Omega^D,\\
    \vc{D}\cdot\vc{n} &= D^N, \qquad &\vc{x}\in\partial\Omega^N,
\end{alignat}
where $\vc{n}$ is the outward-pointing normal vector, and the domain boundary is partitioned such that $\partial\Omega=\partial\Omega^D\cup\partial\Omega^N$ and $\partial\Omega^D\cap\partial\Omega^N=\emptyset$.

Finally, the macroscopic charge density coupling the Vlasov and Poisson equations is obtained by summing up the contributions of all species:
\begin{equation}
    \rho = \sum_\alpha q_\alpha\int_{\mathbb{R}^3} f_\alpha(\vc{x},\vc{v},t)\,d\vc{v}.\label{eq:charge-density}
\end{equation}

\section{Numerical Method}
In the PIC method \cite{birdsallPlasmaPhysicsComputer1985,hockneyComputerSimulationUsing2021}, the continuous distribution function $f_\alpha$ is approximated by a finite set of $N_\alpha$ macroscopic computational particles:
\begin{equation}
    f_\alpha(\vc{x},\vc{v},t) \approx \sum_{p=1}^{N_\alpha} w_p \delta\left(\vc{x}-\vc{x}_p(t)\right)\delta\left(\vc{v}-\vc{v}_p(t)\right),
\end{equation}
where $x_p(t)$ and $v_p(t)$ are the position and velocity of the p-th particle, respectively, and $w_p$ is a weighting factor that represents the number of physical particles combined into one macroscopic simulation particle.
Inserting this discrete representation into the charge density expression yields:
\begin{equation}
    \rho(\vc{x}) = \sum_\alpha q_\alpha\sum_{p=1}^{N_\alpha} w_p \delta(\vc{x}-\vc{x}_p(t)).
\end{equation}
While many PIC codes employ smoothing shape functions for the charge deposition to reduce numerical noise \cite{jacobsHighorderNodalDiscontinuous2006,pfeifferParticleCellSolverBased2019}, all simulations in this work project the delta distribution directly onto the high-order polynomial basis of the HDG-SEM without additional smoothing.

The particle trajectories are governed by the characteristic curves of the Vlasov equation:
\begin{align}
    \frac{\text{d}\vc{x}_p}{\text{d}t} &= \vc{v}_p,\\
    \frac{\text{d}\vc{v}_p}{\text{d}t} &= \frac{q_\alpha}{m_\alpha}\vc{E}(\vc{x}_p).
\end{align}

\subsection{Time Integration}
The particle trajectories are integrated in time using the standard second-order Leapfrog (Störmer-Verlet) scheme \cite{hairerGeometricNumericalIntegration2006}.
By staggering positions and velocities by a half time-step $\Delta t/2$, the method achieves second-order accuracy without the need to solve implicit equations, which makes it computationally efficient for large-scale PIC simulations.
Furthermore, the symplectic nature of the Leapfrog method ensures excellent long-term energy conservation properties.
The Leapfrog scheme is given by the following update equations:
\begin{align}
    \vc{v}_p^{n+1/2}&=\vc{v}_p^{n-1/2} + \Delta t \frac{q_\alpha}{m_\alpha}\vc{E}^n(\vc{x}_p^n),\\
    \vc{x}_p^{n+1}&=\vc{x}_p^n+\Delta t \vc{v}_p^{n+1/2}.
\end{align}

\subsection{Field solver}
In \textsc{PICLas}, the Poisson equation is solved numerically using the hybridizable discontinuous Galerkin (HDG) method.
Originally introduced and analyzed in a series of papers by Cockburn et al. \cite{cockburnSuperconvergentLDGhybridizableGalerkin2008,cockburnHybridizableDiscontinuousGalerkin2009,cockburnDerivationHybridizableDiscontinuous2009,cockburnUnifiedHybridizationDiscontinuous2009}, the method was previously implemented in \textsc{PICLas} to solve the Poisson equation for electrostatic applications using uniform polynomial degrees across all elements~\cite{pfeifferParticleCellSolverBased2019}.
In the present work, the formulation is extended to support element-local p-adaptation.

This section outlines the HDG formulation and its implementation in \textsc{PICLas}, with a focus on p-adaptation for element-local refinement.
The computational domain $\Omega$ is first partitioned into non-overlapping, hexahedral elements $K$.
The boundary $\partial K$ of each element is subdivided into six quadrilateral faces $F_i$, such that $\partial K=\cup_{i=1}^6 F_i$.
The set of all faces for an element is denoted by $\Gamma_K=\{F_1,\dots,F_6\}$.
Each face $F\in\Gamma_K$ is either an inner face shared between two adjacent elements, $F=\partial K^+\cap\partial K^-$, or a boundary face lying on the global Neumann or Dirichlet boundary, $F^{N/D}\in\partial\Omega^{N/D}$.

\subsubsection{Reference element mapping}
For each element $K$, a bijective transformation is defined to map a standard reference element $\hat{K}\in[-1,1]^3$ to the physical element:
\begin{equation}
\vec{x}:\vec{\xi}\in\hat{K}\rightarrow\vec{x}^K(\vec{\xi})\in K,\qquad \vec{\xi}=(\xi,\eta,\zeta)^T.
\end{equation}
Since the mapping is bijective, an inverse mapping $\vec{\xi}^K(\vec{x})$ exists.

The Jacobian matrix of this transformation and its determinant are given by
\begin{equation}
    \bm{\mathcal{J}}^K = \frac{\partial\vec{x}^K}{\partial\vec{\xi}},\qquad J^K=|\bm{\mathcal{J}}^K|
\end{equation}
With this, the divergence operator transforms as $\nabla\cdot \vec{g} = J^{-1}\nabla_{\xi}\cdot(J{\bm{\mathcal{J}}^K}^{-1}\vec{g})$.

In order to transform the integrals over the physical element to integrals over the reference element, the outward-pointing normal vector $\vec{n}$ on the element faces is transformed by
\begin{equation}
    \vec{n} = \frac{J^K{\bm{\mathcal{J}}^K}^{-T}\vec{N}}{J^F},\qquad J^F = \|J^K{\bm{\mathcal{J}}^K}^{-T}\vec{N}\|,
\end{equation}
where $\vec{N}$ is the normal vector on the reference face and $J^F$ is the Jacobian of the transformation from the reference face to the physical face.

Within the reference element, a tensor-product polynomial basis of degree $k$ is defined.
In three dimensions, the scalar basis functions take the form
\begin{equation}
    \bar{\phi}_{ijk}(\xi) = \ell_i(\xi)\ell_j(\eta)\ell_k(\zeta), \qquad i,j,k\in\{0,...,k\},
\end{equation}
and the vector basis functions are given by
\begin{equation}
    \bar{\bm{D}}_{d,ijk}(\xi) = \vec{e}_d\ell_i(\xi)\ell_j(\eta)\ell_k(\zeta), \qquad i,j,k\in\{0,...,k\},\qquad d\in\{1,2,3\},
\end{equation}
where $\vec{e}$ is the standard Cartesian basis vector with $(\vec{e}_d)_i=\delta_{id}$ and $\ell_i$ are Lagrange polynomials with Legendre-Gauss nodes.

The scalar basis functions in element $K$ are then given by $\bar{\phi}_{ijk}^K(\vec{x}) = \bar{\phi}_{ijk}(\vec{\xi}^K(\vec{x}))$.
For the vector basis, the degrees of freedom are chosen as the vector components in physical directions, such that $\bar{\bm{D}}_{d,ijk}^K(\vec{x})={J^K}^{-1}\bm{\mathcal{J}}^K\bar{\bm{D}}_{d,ijk}(\xi^K(\vec{x}))$.

Concretely, the approximate potential and displacement field are given by
\begin{equation}
    \phi(\vec{x})\approx\sum_K\sum_{ijk=0}^{k(K)}\hat{\phi}_{ijk}^K\bar{\phi}_{ijk}^K(\vec{x}),\qquad \bm{D}(\vec{x})\approx\sum_K\sum_{d=1}^3\sum_{ijk=0}^{k(K)}\hat{D}^K_{d,ijk}\bar{\bm{D}}_{d,ijk}^K(\vec{x}).
\end{equation}

\subsubsection{Hybridizable discontinuous Galerkin method}
The weak formulation is obtained by multiplying the Poisson equation by test functions $\bar{\vec{D}}_i^K$ and $\bar{\varphi}_i^K$ and integration over the global domain $\Omega$.
The test functions are chosen from the same set of functions as the basis functions defined above.
\begin{align}
    \int_{\Omega} \frac{1}{\varepsilon}\vec{D}\cdot\bar{\vec{D}}^K_i\,d\vec{x} + \int_{\Omega} \nabla\varphi\cdot\bar{\vec{D}}^K_i\,d\vec{x}&= 0,\\
    \int_{\Omega} \nabla\cdot{\vec{D}}\bar{\varphi}^K_j\,d\vec{x} &= \int_{\Omega} \rho\bar{\varphi}^K_j\,d\vec{x}.
\end{align}

Since the test functions are only non-zero in a single element, the global integration reduces to integrals over the single element $K$.
\begin{align}
    \int_{K} \frac{1}{\varepsilon}\vec{D}\cdot\bar{\vec{D}}_i^K\,d\vec{x} + \int_{K} \nabla\varphi\cdot\bar{\vec{D}}_i^K\,d\vec{x}&= 0,\label{eq:variationalelement1}\\
    \int_{K} \nabla\cdot{\vec{D}}\bar{\varphi}_i^K\,d\vec{x} &= \int_{K} \rho\bar{\varphi}_i^K\,d\vec{x}.\label{eq:variationalelement2}
\end{align}
Integration by parts once for Eq. \eqref{eq:variationalelement1} and twice for Eq. \eqref{eq:variationalelement2} leads to the following system of equations.
\begin{align}
    \int_{K} \frac{1}{\varepsilon}\vec{D}\cdot\bar{\vec{D}}_i^K\,d\vec{x} - \int_{K} \varphi\nabla\cdot\bar{\vec{D}}_i^K\,d\vec{x} + \int_{\partial K} \varphi^*\bar{\vec{D}}_i^K\cdot\vec{n}\,d\vec{x}&= 0,\\
    -\int_{K} \nabla\cdot{\vec{D}}\bar{\varphi}_i^K\,d\vec{x} - \int_{\partial K} \left.(\vec{D}^* - \vec{D})\right|_K \bar{\varphi}_i^K\,d\vec{x} &= \int_{K} \rho\bar{\varphi}_i^K\,d\vec{x}.
\end{align}
In order to connect the elements, a unique trace variable $\lambda$ is introduced, which approximates the electric potential on the element interfaces.

The transformation onto the reference element can also be used to transform each side onto the reference side $(\xi,\eta)\in[-1,1]^2$.
Similarly to the basis functions in the element, the face basis functions are defined as:
\begin{equation}
    \bar{\lambda}^F_{pq}(\vec{x})=\bar{\lambda}_{pq}(\xi^F(\vec{x})),\qquad\bar{\lambda}_{pq}(\xi)=\ell_p(\xi)\ell_q(\eta),\qquad p,q\in{0,\dots,k}.
\end{equation}
With this basis on the side, the potential on the sides is approximated by
\begin{equation}
     \lambda(\vec{x})\approx\sum_F\sum_{pq=0}^{k(F)}\hat{\lambda}_{pq}^F\bar{\lambda}_{pq}^F(\vec{x}).
\end{equation}

As for the elements, the polynomial degree of the solution may change across sides.
Following the choice of Cockburn et. al.~\cite{cockburnUnifiedHybridizationDiscontinuous2009}, we choose for the polynomial degree for the inner side $F$:
\begin{equation}
    k(F) = \max{\left(k(K^+), k(K^-)\right)}, \qquad F = \partial K^+\cap\partial K^-.
\end{equation}
The flux is continuous across element faces.
Projecting this condition onto the newly introduced face basis functions yields the following equations for inner faces $F = \partial K^+\cap\partial K^-$ and Neumann faces $F^N\in\partial K\cap\partial\Omega^N$:
\begin{align}
    \int_F\,\left((\vec{D}^*\cdot\vec{n})|_{K^+} + (\vec{D}^*\cdot\vec{n})|_{K^-}\right)\bar{\lambda}_i^F\,d\vec{x} = 0,\\
    \int_{F^N}\,\left((\vec{D}^*\cdot\vec{n})|_{K} - D_F^N\right)\bar{\lambda}_i^{F^N}\,d\vec{x} &= 0.
\end{align}

What remains is to find adequate choices for the numerical traces to close the system.
Following \cite{cockburnUnifiedHybridizationDiscontinuous2009}, we choose on the interface of each element $K$:
\begin{alignat}{3}
    (\vec{D}^*\cdot\vec{n})|_K &= (\vec{D}\cdot\vec{n})|_K + \tau(\varphi - \varphi^*),\qquad &\vec{x}&\in\partial K,\\
    \varphi^* &= \lambda, \qquad &\vec{x}&\in\partial K \backslash \partial\Omega^D,\\
    \varphi^* &= \lambda^D, \qquad &\vec{x}&\in\partial K \cap \partial\Omega^D,
\end{alignat}
with a stabilization parameter $\tau>0$.
The stabilization parameter defines to what extent a discontinuity of the potential on the interfaces is allowed.
For instance, as $\tau\to\infty$, the system converges to the continuous Galerkin solution with a continuous electric potential.
For the value of the stabilization parameter, we choose $\tau \propto |K|^{-1/3}$ with $|K|$ being the volume of the element as suggested by Cockburn et al. \cite{cockburnSuperconvergentLDGhybridizableGalerkin2008} for optimal convergence properties.

Inserting the numerical traces in the equation for an element $K$ yields
\begin{multline}
    \int_{K} \frac{1}{\varepsilon}\vec{D}\cdot\bar{\vec{D}}_i^K\,d\vec{x} - \int_{K} \varphi\nabla\cdot\bar{\vec{D}}_i^K\,d\vec{x} + \int_{\partial K\backslash\partial\Omega^D} \lambda\bar{\vec{D}}_i^K\cdot\vec{n}\,d\vec{x}= \\-\int_{\partial K\cap\partial\Omega^D} \lambda\bar{\vec{D}}_i^K\cdot\vec{n}\,d\vec{x},
\end{multline}
\begin{multline}
    -\int_{K} \nabla\cdot\vec{D}\bar{\varphi}_i^K\,d\vec{x} - \int_{\partial K} \tau\varphi \bar{\varphi}_i^K\,d\vec{x} + \int_{\partial K\backslash\Omega^D} \tau\lambda \bar{\varphi}_i^K\,d\vec{x} = \\-\int_{\partial K\cap\Omega^D} \tau\lambda \bar{\varphi}_i^K\,d\vec{x} + \int_{K} \rho\bar{\varphi}_i^K\,d\vec{x}
\end{multline}
and for a shared face $F=\partial K^-\cap\partial K^+$
\begin{equation}
    \sum_{K\in\{K^+,K^-\}}\int_F\,(\vec{D}\cdot\vec{n})|_{K} \bar{\lambda}_i^F\,d\vec{x} + \int_F\,(\tau\varphi)|_{K} \bar{\lambda}_i^F\,d\vec{x} - \int_F\,(\tau\lambda)|_{K} \bar{\lambda}_i^F\,d\vec{x} = 0.
\end{equation}
For Neumann boundaries, the contribution of one element is simply replaced by the given field.
Thus, we will ignore Neumann boundaries for now on, as the procedure can be written analogously to inner sides.

Before we evaluate the integrals, the equations are transformed into reference coordinates.
For the element $K$, this leads to
\begin{multline}
    \int_{\hat{K}} \frac{1}{\varepsilon}\frac{1}{J^K}{\mathcal{J}^K}^{-1}\vec{D}\cdot{\mathcal{J}^K}^{-1}\bar{\vec{D}}_i\,d\vec{\xi} - \int_{\hat{K}} \varphi\nabla_{\xi}\cdot\bar{\vec{D}}_i\,d\vec{\xi} + \int_{\partial \hat{K}\backslash\Omega^D} \lambda\bar{\vec{D}}_i\cdot\vec{N}\,d\vec{\xi} = \\
    -\int_{\partial \hat{K}\cap\Omega^D} \lambda^D\bar{\vec{D}}_j\cdot\vec{N}\,d\vec{\xi},\label{eq:hdgref1}
\end{multline}
\begin{multline}
    -\int_{\hat{K}} \nabla_\xi\cdot{\vec{D}}\bar{\varphi}_i\,d\vec{\xi} - \int_{\partial \hat{K}} J^K\tau\varphi\bar{\varphi}_i\,d\vec{\xi} + \int_{\partial \hat{K}\backslash \Omega^D}J^K\tau\lambda\bar{\varphi}_j\,d\vec{\xi} =
    \\-\int_{\partial \hat{K}\cap \Omega^D}J^K\tau\lambda^D\bar{\varphi}_j\,d\vec{\xi} + \int_{\hat{K}} J^K\rho\bar{\varphi}_j\,d\vec{\xi}.\label{eq:hdgref2}
\end{multline}
and for $F=\partial K^-\cap\partial K^+$ to
\begin{equation}
    \sum_{K\in\{K^+,K^-\}}\left(\int_{\hat{F}} (\vec{D}\cdot\vec{N})|_K \bar{\lambda}_i\,d\vec{\xi} + \int_{\hat{F}} (J\tau\varphi)|_K \bar{\lambda}_i\,d\vec{\xi} - \int_{\hat{F}} (J\tau)|_K \lambda\bar{\lambda}_i\,d\vec{\xi}\right) = 0.\label{eq:hdgref3}
\end{equation}

\subsubsection{Approximation of the integrals}
All integrals on the reference element and side are approximated using Gaussian quadrature.
By using the same integration points as nodes for the basis functions, the integration can be performed efficiently as many summands vanish due to the Kronecker-$\delta$ property of the basis functions.
This in turn leads to sparse matrices that are needed to solve the system.
For instance, the first integral in Eq. \eqref{eq:hdgref1} reduces to a block-diagonal form, where each block of size $3\times 3$ corresponds to a node in the element:
\begin{equation}
    \int_{\hat{K}} \frac{1}{\varepsilon}\frac{1}{J^K}{\mt{\mathcal{J}}^K}^{-1}\bar{\vc{D}}_{lmn}\cdot{\mt{\mathcal{J}}^K}^{-1}\bar{\vec{D}}_{ijk}\,d\vec{\xi} \approx \frac{1}{\varepsilon}\frac{1}{J^K_{ijk}}{\mt{\mathcal{J}}^{K}_{ijk}}^{-T}{\mt{\mathcal{J}}^{K}_{ijk}}^{-1} \,\omega_i\omega_j\omega_k\delta_{il}\delta_{jm}\delta_{kn} \eqcolon \mt{\mathcal{A}}_{ijk,lmn}.
\end{equation}
Here, $\omega_i$ denote the weights of the Gaussian quadrature and $J^K_{ijk}$ and $\mt{\mathcal{J}}^K_{ijk}$ are the values of the Jacobian determinant and the Jacobi matrix at the node $(\xi_i,\eta_j,\zeta_k)$.

Gaussian quadrature is exact for polynomials of degree up to $2k-1$.
This means that an error when calculating the integrals may be introduced if the Jacobi determinant $J$ or the inverse Jacobi matrix $\mathcal{J}^{-1}$ is not constant.

Since the polynomial degree can vary in each element and on each side, the accuracy of the numerical quadrature does also vary.
However, since the polynomial degree on a side is always larger or equal to the polynomial degree of the adjacent elements, no new integration error will be introduced here.

\subsubsection{Construction of the global system}
Eqs. \eqref{eq:hdgref1} and \eqref{eq:hdgref2} can be evaluated for all basis functions.
Doing this, the resulting equations in each element $K$ can be written as a linear system:
\begin{equation}
    \begin{pmatrix}
        \mt{\mathcal{A}} & \mt{\mathcal{B}}^T\\
        \mt{\mathcal{B}} & \mt{\mathcal{D}}
    \end{pmatrix}_K
    \begin{pmatrix}
        \vc{\hat{D}}\\ \vc{\vc{\hat{\phi}}}
    \end{pmatrix}_K
    +
    \begin{pmatrix}
        \mt{\mathcal{C}}^T\\ \mt{\mathcal{E}}^T
    \end{pmatrix}_{\partial K\backslash\partial\Omega^D}
    \vc{\hat{\lambda}}_{\partial K\backslash\partial\Omega^D}
    =
    -\begin{pmatrix}
        \mt{\mathcal{C}}^T\\
        \mt{\mathcal{E}}^T
    \end{pmatrix}_{\partial K\cap\partial\Omega^D}
    \vc{\hat{\lambda}}_{\partial K\cap\partial\Omega^D}
    +\begin{pmatrix}
        \vc{0}\\
        \vc{\rho}
    \end{pmatrix}_K
    \label{eq:element-system}
\end{equation}
Here, The vectors $\vc{\hat{D}}$ and $\vc{\hat{\phi}}$ contain the unknown degrees of freedom in the element.

The vector $\vc{\hat{\lambda}}_{\partial K\backslash\partial\Omega^D}$ contains the degrees of freedom on all inner and Neumann sides of the element:
\begin{equation}
    \vc{\hat{\lambda}}_{\partial K\backslash\partial\Omega^D} =
    [\vc{\hat{\lambda}}_F]_{F\in\Gamma_K,F\nsubseteq\partial\Omega^D}
\end{equation}
The two matrices $\mt{\mathcal{C}}_{\partial K\backslash\partial\Omega^D}$ and $\mt{\mathcal{E}}\_{\partial K\backslash\partial\Omega^D}$ that connect the element to the sides can be split into contributions to each side in a similar manner:
\begin{align}
    \mt{\mathcal{C}}_{\partial K\backslash\partial\Omega^D} &= [\mt{\mathcal{C}}_{F,K}]_{F\in\Gamma_K,F\nsubseteq\partial\Omega^D}\\
    \mt{\mathcal{E}}_{\partial K\backslash\partial\Omega^D} &= [\mt{\mathcal{E}}_{F,K}]_{F\in\Gamma_K,F\nsubseteq\partial\Omega^D}.
\end{align}

Evaluating Eq. \eqref{eq:hdgref3} that connects the elements leads to
\begin{equation}
    \sum_{K\in\{K^+,K^-\}}{
        \begin{pmatrix}
            \mt{\mathcal{C}}_{F,K} & \mt{\mathcal{E}}_{F,K} & \mt{\mathcal{F}}_{F}
        \end{pmatrix}
        \begin{pmatrix}
            \vc{\hat{D}}_K \\
            \vc{\hat{\phi}}_K \\
            \vc{\hat{\lambda}}_F
        \end{pmatrix} = 0
        }
        \label{eq:transmission_system}
\end{equation}
for an inner side $F = \partial K^+\cap\partial K^-$ with adjacent elements $K^+$ and $K^-$.
By solving the system \eqref{eq:element-system} for the unknowns $\vc{D}_K$ and $\vc{\phi}_K$ and substituting them into Eq.~\eqref{eq:transmission_system}, the element-interior degrees of freedom are eliminated.
This yields a global system exclusively in terms of the trace unknowns:
\begin{equation}
    \sum_{K\in{K^+,K^-}}\sum_{\substack{F'\in\Gamma_K\\F'\nsubseteq\partial\Omega^D}}{\mt{\mathcal{S}}_{F,F'}\vc{\hat{\lambda}}_{F'}} = \vc{r}_F
\end{equation}
where
\begin{align}
    \mt{\mathcal{S}}_{F,F'} &= \mt{\tilde{\mathcal{E}}}_{\Gamma,K}\mt{\tilde{\mathcal{D}}}_K\mt{\tilde{\mathcal{E}}}_{F',K}^T + \mt{\mathcal{C}}_{F,K}\mt{\mathcal{A}}_K^{-1}\mt{\mathcal{C}}_{F',K}^T - \delta_{F,F'}\mt{\mathcal{F}}_F, \label{eq:system_rhs}\\
    \bm{r}_F&=\sum_{K\in\{K^+,K^-\}}{\left(\mt{\tilde{\mathcal{E}}}_{F,K}\mt{\tilde{\mathcal{D}}}_K^{-1}\vc{\rho}_K - \sum_{\substack{F'\in\Gamma_K\\F'\subseteq\partial\Omega^D}}\mt{\mathcal{S}}_{F,F'}\vc{\hat{\lambda}}^D_{F'} \right)}
    \label{eq:f}
\end{align}
and
\begin{equation}
    \mt{\tilde{\mathcal{E}}}_{F,K} = \mt{\mathcal{E}}_{F,K} - \mt{\mathcal{C}}_{F,K}\mt{\mathcal{A}}_K^{-1}\mt{\mathcal{B}}_K^T,\qquad
    \mt{\tilde{\mathcal{D}}}_{K} = \mt{\mathcal{D}}_{K} - \mt{\mathcal{B}}_{K}\mt{\mathcal{A}}_K^{-1}\mt{\mathcal{B}}_K^T.
\end{equation}
Neumann faces are treated analogously.

Collecting the equations for all sides results in the global coupled system
\begin{equation}
    \mt{\mathcal{S}}\vc{\hat{\lambda}}=\vc{r}
\end{equation}
that needs to be solved.

Finally, after solving the potential on the side, the electric potential and electric field are computed element-wise by inverting Eq.~\eqref{eq:element-system}:
\begin{align}
    \vc{\hat{\phi}}_K &= \mt{\tilde{\mathcal{D}}}^{-1}_K \vc{\rho}_K - \mt{\tilde{\mathcal{D}}}^{-1}_K\mt{\tilde{\mathcal{E}}}^T_{\partial K} \vc{\lambda}_{\partial K} \label{eq:postphi}\\
    \vc{\hat{D}}_K &= -\mt{\mathcal{A}}_K^{-1} \mt{\mathcal{B}}^T_K \vc{\hat{\phi}}_K - \mt{\mathcal{A}}^{-1}_K \mt{\mathcal{C}}^T_{\partial K} \vc{\lambda}_{\partial K}.\label{eq:postD}
\end{align}

\section{Implementation}
Allowing for variable polynomial degrees across different elements and their corresponding faces requires a highly flexible data structure.
Unlike uniform high-order methods, where the solution can be stored in dense, multi-dimensional arrays, the p-adaptive implementation in \textsc{PICLas} utilizes derived data types.
Specifically, distinct element and face types are defined, with each instance of these types containing the solution based on its locally assigned polynomial degree.
These instances are then organized into one-dimensional arrays.

In order to connect the individual elements and to construct the global system matrix, a topological mapping is required to associate local face coordinates with the adjacent elements.
For 3D hexahedral elements, a shared quadrilateral face can be oriented in five different ways relative to the local reference coordinate system of the adjacent elements.
To resolve this, \textsc{PICLas} stores the global side index and its rotation relative to the element for each local face of an element, which allows for the correct assembly of the global system matrix and ensures that the solution is properly connected across element interfaces.

To optimize the computational performance, the matrices $\mt{\hat{\mathcal{E}}}_K$ and $\mt{\hat{\mathcal{D}}}_K$ are precomputed and stored for each element.
These matrices are required to populate the right-hand side of the global system in Eq. \eqref{eq:system_rhs} and to reconstruct the element-interior electric potential and displacement fields from the surface solution using Eqs. \eqref{eq:postphi} and \eqref{eq:postD}.
The remaining matrices can be efficiently computed on-the-fly to minimize the memory footprint.

The final global linear system for the interface potentials is assembled and solved using the parallel numerical library \textsc{PETSc} \cite{petsc-web-page}.
To facilitate this, a unique global index is assigned to each integration node on the inner and Neumann boundary faces.
For smaller systems, or when sufficient computational resources are available, a direct Cholesky decomposition is used.
For large-scale systems where direct inversion is not feasible, an iterative Conjugate Gradient solver is employed, accelerated by the algebraic multigrid preconditioner \textit{BoomerAMG} provided by the \textsc{HYPRE} library \cite{hypre} within \textsc{PETSc}.

\subsection{P-adaptation strategy}
To ensure an efficient p-adaptive method, the optimal polynomial degree within each element must be automatically adapted based on the local solution.
In this work, an iterative modal analysis strategy is employed to determine the required polynomial degree between simulations.
This approach is closely related to the strategy proposed by Mavriplis \cite{mavriplis1994adaptive} for adaptation in the context of spectral element methods for fluid dynamics.

In spectral methods, it is well established that the modal contributions of a sufficiently smooth solution decay exponentially with respect to the mode number \cite{boyd2001chebyshev}.
Thus, by expanding local physical quantities such as the electric potential $\varphi$ and the charge densities $\rho$ of each species into the orthogonal Legendre basis, the required polynomial degree can be estimated.
For instance, denoting the Legendre polynomials as $\Psi_k(\xi)$, the local charge density within a 1D reference element can be expressed as
\begin{equation}
    \rho(\xi) = \sum_{k=0}^\infty \hat{\rho}_k \Psi_k(\xi).
\end{equation}
For a noise-free, smoothly varying field, the required polynomial degree can be estimated by identifying the lowest degree $k$ for which the squared normalized modal contribution falls below a predefined threshold $\varepsilon$.

However, utilizing the charge density $\rho$ as a refinement indicator introduces a significant challenge due to the inherent statistical noise associated with the particle representation in PIC simulations.
Applying a simple constant threshold $\varepsilon$ would lead to excessive refinement in regions with high particle noise, effectively attempting to resolve statistical fluctuations rather than the physical solution.
To decouple the noise from the physical solution, we introduce a dynamic threshold that accounts for the noise level in the charge density.
For each element and for each Legendre mode $k$, we estimate the variance $\sigma_k^2$ of the modal contribution caused by particle sampling, as detailed in Appendix \ref{app:threshold}.
If the squared normalized contribution of a mode is small compared to the variance, i.e. $\hat{\rho}_k^2 < \varepsilon\sigma_k^2$, we can assume that the contribution of this mode is dominated by noise and contains no resolvable physical signal, and thus we truncate the expansion at this mode.

\section{Validation results}
In this section, the implementation of the p-adaptive HDG-SEM in \textsc{PICLas} is validated and compared against the non-adaptive uniform formulation.
The accuracy of the solver is evaluated by analyzing the normalized $L^2$ error of the electric potential relative to known analytical solutions, defined over the computational domain $\Omega$ as
\begin{equation}
    \| \varphi \|_{L^2(\Omega)} = \sqrt{\frac{1}{|\Omega|}\int_\Omega \varphi^2(\vec{x})\,d\vec{x}} = \sqrt{\frac{\sum_K \int_{\hat{K}} \varphi^2(\vec{x}^K(\vec{\xi}))J\,d\vec{\xi}}{\sum_K{|K|}}}.
\end{equation}
The integrations are performed using Gaussian quadrature on the reference element.
All computational meshes utilized in the following test cases were generated using the high-order preprocessor \textsc{pyHOPE} \cite{Kopper2025}.

\subsection{Dielectric sphere}
In order to validate the p-adaptive HDG-SEM formulation without particles, we first consider an analytical benchmark featuring a dielectric sphere.
The sphere, characterized by its radius $R$ and relative permittivity $\varepsilon_r$, is placed in a uniform electric field $\vc{E} = (0,0,E_z)^T$.
The analytical solution for the electric potential is given by \cite{jackson_classical_1999}:
\begin{equation}
    \varphi(\vc{x}) = \begin{dcases}
        -\frac{3}{\varepsilon_r+2}E_z z, & \text{if } |\vc{x}| > R,\\
        \left(\frac{\varepsilon_r-1}{\varepsilon_r+2}\frac{R^3}{\|\vc{x}\|^3}-1\right)E_z z, & \text{if } |\vc{x}| \leq R.
    \end{dcases}
\end{equation}

\begin{figure}[htbp]
    \centering
    \begin{subfigure}[b]{0.4\textwidth}
        \includegraphics[width=\textwidth]{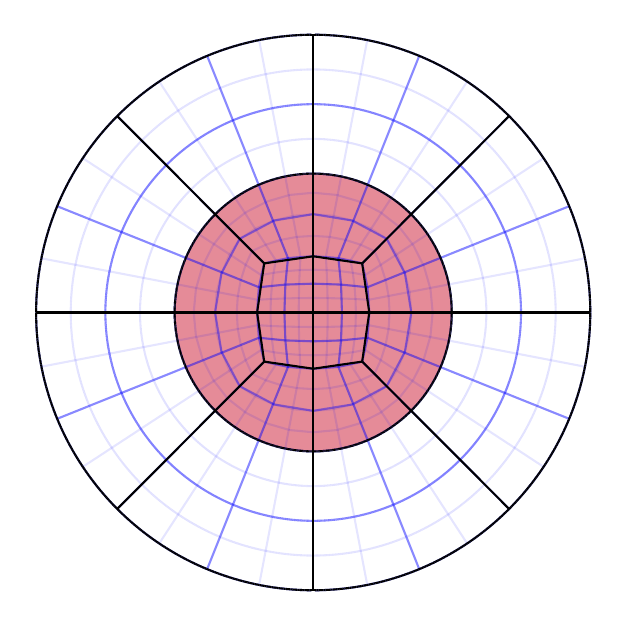}
        \caption{Different refinement levels used for the convergence tests.}
        \label{fig:refinement-levels}
    \end{subfigure}
    \hfill
    \begin{subfigure}[b]{0.4\textwidth}
        \includegraphics[width=\textwidth]{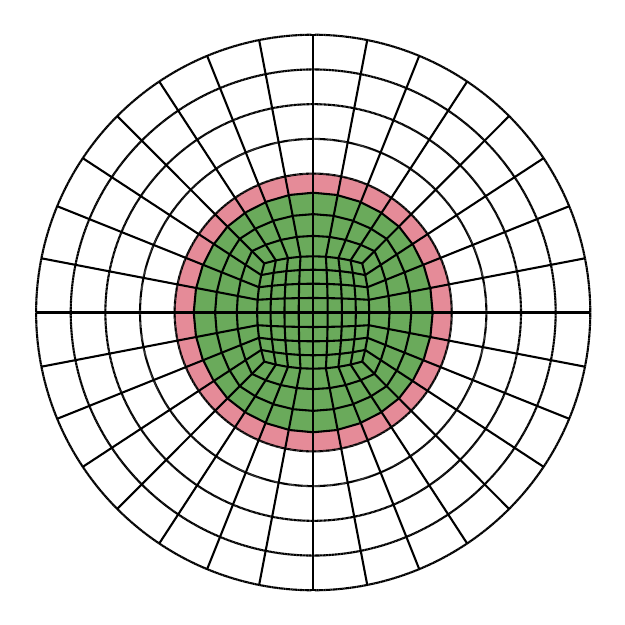}
        \caption{Area inside the dielectric sphere where P-adaptation is used, shown in green.}
        \label{fig:p-adaptation-area}
    \end{subfigure}
    \caption{Refinement levels and P-adaptation area used for the dielectric sphere convergence tests. The dielectric sphere is shown in red.}
\end{figure}

The curved computational domain is defined as a sphere of radius $R_{\text{mesh}} = 2R$.
To accurately conform to the dielectric interface, curved elements are employed at the surface of the sphere, while the interior is discretized using trilinear elements, as shown in Figure \ref{fig:refinement-levels}.
The domain boundaries are treated as Dirichlet boundaries, with the potential set to the exact analytical solution.

\begin{figure}[htbp]
    \centering
    \begin{tikzpicture}
  \begin{axis}[
    hide axis,
    axis equal image,
    view={0}{0},
    xmin=0,
    xmax=659,
    ymin=0,
    ymax=654,
    colorbar right,
    colormap={rgb}{rgb255=(81,87,110) rgb255=(0,0,255) rgb255=(0,255,255) rgb255=(0,255,0) rgb255=(255,255,0) rgb255=(255,0,0) rgb255=(224,0,255)},
    point meta min=0.25, point meta max=2.5,
    colorbar style={
      ytick={0.5, 1.0, 1.5, 2.0, 2.5},
      title={$|\bm{E}| / \SI{}{\volt}$}, title style={},
      major tick length=2pt,
    },
  ]
    \addplot graphics [xmin=0,xmax=659,ymin=0,ymax=654]{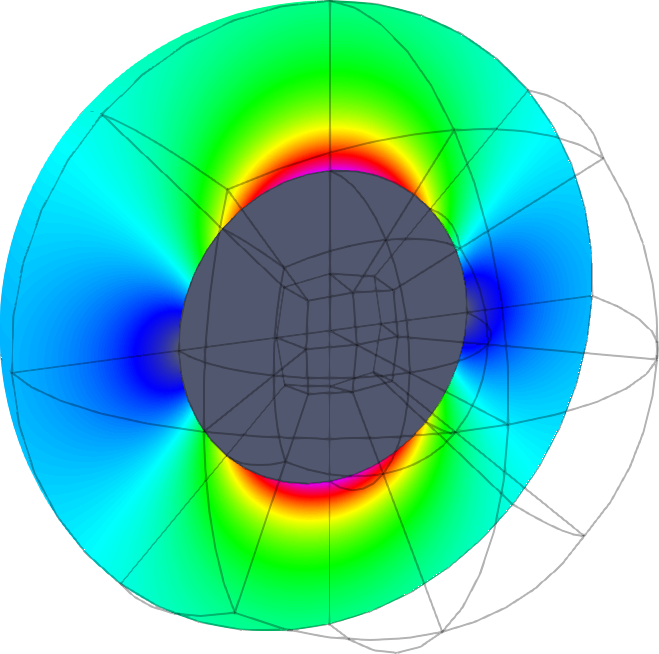};
  \end{axis}
\end{tikzpicture}
    \caption{Cutaway showing the simulated absolute value of the electric field and the edges of the curved mesh.}
    \label{fig:diel-sphere-cutaway}
\end{figure}

Figure \ref{fig:diel-sphere-cutaway} shows a cutaway of the simulated absolute value of the electric field.
Inside the sphere, the electric field is constant and in turn, the electric potential is linear.
Consequently, a low polynomial degree is sufficient to represent the solution inside the sphere.
While an exact integration of the system matrices would only require $N=1$ to perfectly capture this linear potential, the trilinear mapping of the non-cuboid interior elements introduces an integration error due to the non-constant Jacobian.
To prevent integration errors induced by this geometric mapping, the p-adapted polynomial degree inside the sphere is set to $N=2$, as shown in Figure \ref{fig:p-adaptation-area}.

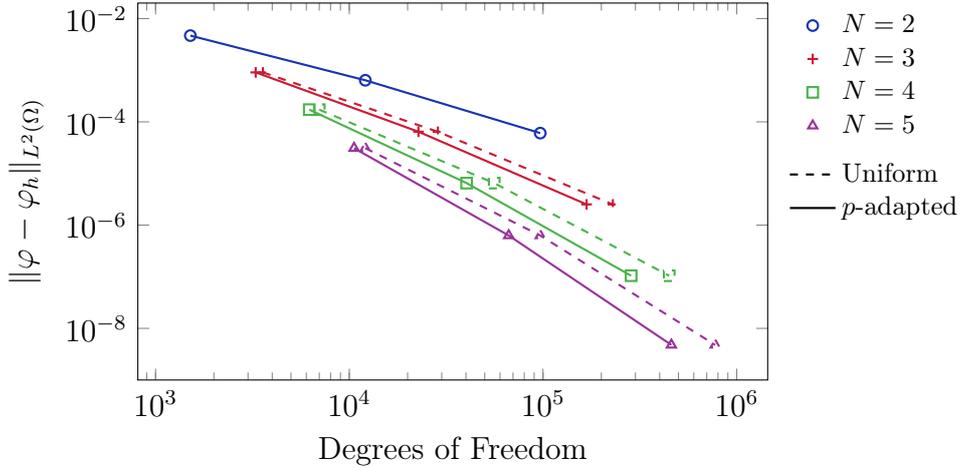
\begin{figure}[htbp]
    \centering
    \begin{tikzpicture}
  \begin{axis}[
      width=0.6\textwidth,
      height=0.4\textwidth,
      xmode=log,
      ymode=log,
      xlabel={Degrees of Freedom},
      ylabel={$ \lVert \varphi-\varphi_h \rVert_{L^2(\Omega)} $},
      ymin=1e-9, ymax=2e-2,
      ytick={1e-8, 1e-6, 1e-4, 1e-2},
      yticklabels={$10^{-8}$, $10^{-6}$, $10^{-4}$, $10^{-2}$},
      scaled ticks=false,
      legend style={
        at={(1.02,1)}, anchor=north west,
        draw=none, fill=none, font=\footnotesize,
      },
      legend cell align=left,
      legend columns=1,
    ]
    \addplot[mark=o       , color=col1, thick, solid, forget plot] table[x=nDOFs, y=L2, col sep=comma]{data/dielectric-sphere/norms-N2-2.csv};
    \addplot[mark=+       , color=col2, thick, dashed, forget plot] table[x=nDOFs, y=L2, col sep=comma]{data/dielectric-sphere/norms-N3-3.csv};
    \addplot[mark=square  , color=col3, thick, dashed, forget plot] table[x=nDOFs, y=L2, col sep=comma]{data/dielectric-sphere/norms-N4-4.csv};
    \addplot[mark=triangle, color=col4, thick, dashed, forget plot] table[x=nDOFs, y=L2, col sep=comma]{data/dielectric-sphere/norms-N5-5.csv};

    \addplot[mark=+       , color=col2, thick, solid, forget plot] table[x=nDOFs, y=L2, col sep=comma]{data/dielectric-sphere/norms-N3-2.csv};
    \addplot[mark=square  , color=col3, thick, solid, forget plot] table[x=nDOFs, y=L2, col sep=comma]{data/dielectric-sphere/norms-N4-2.csv};
    \addplot[mark=triangle, color=col4, thick, solid, forget plot] table[x=nDOFs, y=L2, col sep=comma]{data/dielectric-sphere/norms-N5-2.csv};

    \addlegendimage{mark=o, color=col1, thick, only marks}
    \addlegendentry{$N=2$}
    \addlegendimage{mark=+, color=col2, thick, only marks}
    \addlegendentry{$N=3$}
    \addlegendimage{mark=square, color=col3, thick, only marks}
    \addlegendentry{$N=4$}
    \addlegendimage{mark=triangle, color=col4, thick, only marks}
    \addlegendentry{$N=5$}

    \addlegendimage{empty legend}
    \addlegendentry{}

    \addlegendimage{draw=black, thick, dashed, no marks}
    \addlegendentry{Uniform}
    \addlegendimage{draw=black, thick, solid, no marks}
    \addlegendentry{$p$-adapted}

  \end{axis}
\end{tikzpicture}
    \caption{Convergence of the normalized $L^2$ error of the electric potential for the dielectric sphere test case with different polynomial degrees. The results with (solid) and without (dashed) p-adaptation are compared at different polynomial degrees. In the P-adaptive case, the polynomial degree for the elements inside the dielectric sphere and not adjacent to its boundary is set to $2$.}
    \label{fig:dielectric-sphere-convergence-single}
\end{figure}

Convergence studies were were conducted with varying polynomial degrees between $N=2$ and $N=5$ for both the p-adaptive and non-adaptive cases on three different refinement levels, depicted in Figure \ref{fig:refinement-levels}.
As demonstrated in the $L^2$ error convergence plots in Figure \ref{fig:dielectric-sphere-convergence-single}, both the uniform and p-adaptive formulations exhibit the expected order of convergence.
The p-adapted simulations achieve nearly identical accuracy to the non-adaptive simulations, while significantly reducing the number of degrees of freedom by using a lower polynomial degree in the interior of the sphere.
As the mesh is refined and the total number of elements increases, the relative number of low-order elements in the p-adaptive case also increases, leading to a more pronounced shift in the convergence plot compared to the non-adaptive case.

\subsection{One-dimensional plasma sheath}
The solver was next validated in a fully coupled electrostatic PIC simulation of a one-dimensional plasma sheath.
When a collisionless plasma interacts with a perfectly conducting charged wall, localized charge separation occurs, establishing a sheath characterized by strong electric field gradients.
The exact solution for the normalized electric potential $\chi = -q_\text{e} \varphi / (k_\text{B}T_\text{e})$ within the plasma sheath is given by \cite{kuriemannBohmCriterionSheath1991}
\begin{equation}
    \frac{1}{2}\left(\frac{\partial\chi}{\partial\xi}\right)^{2} = \vartheta^2\left(\sqrt{1+\frac{2\chi}{\vartheta^2}}-1\right) + \exp{(-\chi)}-1,
\end{equation}
where $\xi=x/\lambda_D$ is the spatial coordinate normalized by the Debye length, and $\vartheta = v_I/\sqrt{k_\text{B}T_e/m_I}$ is the dimensionless velocity of the ions entering the sheath.

The computational domain has a length of $L=\SI{0.03}{\meter}$.
The macroscopic properties at the open inflow boundary (left) model a quasi-neutral plasma consisting of electrons and Hydrogen ions, with their respective parameters and values given in Table \ref{tab:plasma-sheath-species}.
A particle weighting of $\omega_p=10^6$ was chosen, resulting in approximately $\SI{3e4}{}$ particles per species.
At the solid wall boundary (right), ions are absorbed while electrons are reflected specularly.
For the field solver, Dirichlet boundary conditions are applied with fixed potentials, $\varphi(0)=\SI{0}{\volt}$ and $\varphi(L)=\SI{-0.18011}{\volt}$.
A time step of $\Delta t=\SI{1E-8}{\second}$ ensures full resolution of the electron dynamics.

\begin{table}
  \caption{Species-specific parameters and inflow conditions used for the different plasma sheath simulations.}\label{tab:plasma-sheath-species}
  \begin{center}
    \begin{tabular}{llrr}
      \hline\rule{0pt}{12pt}
      & & Electrons & H Ions\\
      \hline\rule{-4pt}{12pt}
      Mass & $m \,/\, \SI{}{\kg}$ & $\SI{9.109E-31}{}$ & $\SI{1.673E-27}{}$\\
      Charge & $q \,/\, \SI{}{\coulomb}$ & $\SI{-1.602E-19}{}$ & $\SI{1.602E-19}{}$\\
      Number density & $n \,/\, \SI{}{\per\meter\cubed}$ & $\SI{1E12}{}$ & $\SI{1E12}{}$\\
      Temperature & $T \,/\, \SI{}{\kelvin}$ & $\SI{1000}{}$ & $\SI{1000}{}$\\
      Velocity & $v \,/\, \SI{}{\meter\per\second}$ & $\SI{0}{}$ & $\SI{11492.19}{}$\\[2pt]
      \hline
    \end{tabular}
  \end{center}
\end{table}

\begin{table}
  \caption{Spatial discretization parameters used for the different plasma sheath simulations. In the last row, the relative $L^2$ error of the simulated potential is shown.}\label{tab:plasma-sheath}
  \begin{center}
    \begin{tabular}{llrrrr}
      \hline\rule{0pt}{12pt}
      & & Case 1 & Case 2 & Case 3 & Case 4\\
      \hline\rule{-4pt}{12pt}
      Number of elements & $N_e$ & $4$ & $4$ & $4$ & $32$\\
      Polynomial degree & $N$ & $1$ & $4$ & $1,4$ & $1$\\
      Number of DOFs & & 8 & 20 & 14 & 64\\
      $||\varphi-\varphi_h||_{L^2(\Omega)}/||\varphi||_{L^2(\Omega)}$ & & $\SI{2.775E-1}{}$ & $\SI{5.308E-3}{}$ & $\SI{4.520E-3}{}$ & $\SI{6.106E-3}{}$\\[2pt]
      \hline
    \end{tabular}
  \end{center}
\end{table}

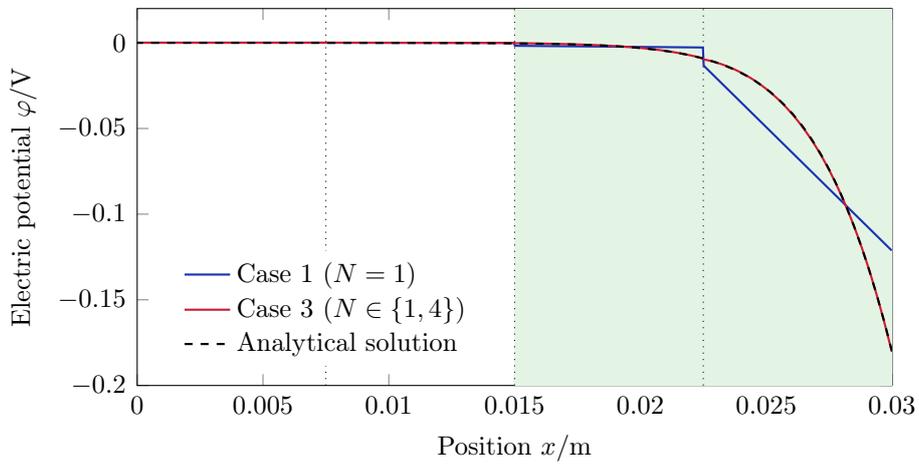
\begin{figure}
    \centering
    \begin{tikzpicture}
    \begin{axis}[
      scaled ticks=false,
      width=0.7\textwidth,
      height=0.4\textwidth,
      xlabel={Position $x / \text{m}$},
      xticklabel style={
      /pgf/number format/fixed,
      /pgf/number format/precision=3
      },
      xmin=0,
      xmax=0.03,
      xtick={0, 0.005, 0.01, 0.015, 0.02, 0.025, 0.03},
      yticklabel style={
      /pgf/number format/fixed,
      /pgf/number format/precision=2
      },
      ylabel={Electric potential $\varphi / \text{V}$},
      ymin=-0.2,
      ymax=0.02,
      cycle list name=exotic,
      font=\footnotesize,
      unbounded coords=jump,
      legend style={at={(0.05,0.05)},anchor=south west,draw=none},
      legend cell align={left},
      ]

      \filldraw[fill=col3!15, draw=none] (0.015,-0.2) rectangle (0.03,0.02);

      \draw[black,dotted] (0.0075,-0.2) -- (0.0075,0.02);
      \draw[black,dotted] (0.015,-0.2) -- (0.015,0.02);
      \draw[black,dotted] (0.0225,-0.2) -- (0.0225,0.02);

      \addplot[mark=none,color=col1,thick] table[x=x,y=phi,col sep=comma] {data/plasma-sheath/4-1-1_curve.csv}; \addlegendentry{Case 1 ($N=1$)}
      \addplot[mark=none,color=col2,thick] table[x=x,y=phi,col sep=comma] {data/plasma-sheath/4-4-1_curve.csv}; \addlegendentry{Case 3 ($N\in\{1,4\}$)}

      \addplot[mark=none,color=black,dashed,thick] table[x=X,y=Phi_ana, col sep=comma] {data/plasma-sheath/analytical_correct.txt}; \addlegendentry{Analytical solution}

    \end{axis}
  \end{tikzpicture}
    \caption{Electric potential for two plasma sheath simulations on a mesh with four elements. In the first simulation, a low polynomial degree ($N=1$) is chosen. In the second simulation, the polynomial degree is set to $N=4$ in the elements marked in green. The analytical solution is shown in black, as a dashed line.}
    \label{fig:plasma-sheath}
\end{figure}

To evaluate the numerical efficiency of the p-adaptive method, four distinct spatial discretization cases were tested with varying numbers of elements and polynomial degrees, as summarized in Table \ref{tab:plasma-sheath}.
In Case 1, a coarse, uniform low-order discretization was used, which fails to resolve the sharp potential gradient within the sheath.
Case 2 uniformly increases the polynomial degree to $N=4$ across all elements, achieving excellent agreement with the analytical solution with only four elements.
Case 3 employs the p-adaptive approach: $N=4$ is specified in the two elements adjacent to the wall, while the remaining elements use a lower polynomial degree of $N=1$.

Comparing Case 2 and Case 3 highlights the advantage of the p-adaptive method, which maintains the high precision of the uniform high-order simulation while reducing the global degrees of freedom for the field solution.
Finally, Case 4 demonstrates standard h-refinement with a uniform low-order approximation ($N=1$) across all elements.
In this case, the number of elements was increased to 32 in order to achieve a similar $L^2$ error as the high-order simulations.
Compared to the p-adapted Case 3, more than double the degrees of freedom were required to obtain a result with similar accuracy, which results in a less efficient simulation and greater memory demand.

\subsection{Ion optic}
Finally, we consider a benchmark problem based on the geometry and operating conditions of a gridded ion thruster, similar to configurations investigated in previous works \cite{binderDevelopmentApplicationPICLas2019,pfeifferParticleCellSolverBased2019}.
To isolate and evaluate the performance of the p-adaptive HDG-SEM under strong-localized electric fields while maintaining a manageable computational cost, a simplified two-dimensional axisymmetric model of a single ion optic aperture is employed.
This test case serves a dual purpose: it validates the solver's capability to handle plasma simulations in an axisymmetric 2D setting, and it demonstrates the advantages of p-adaptation in terms of accuracy and computational efficiency.

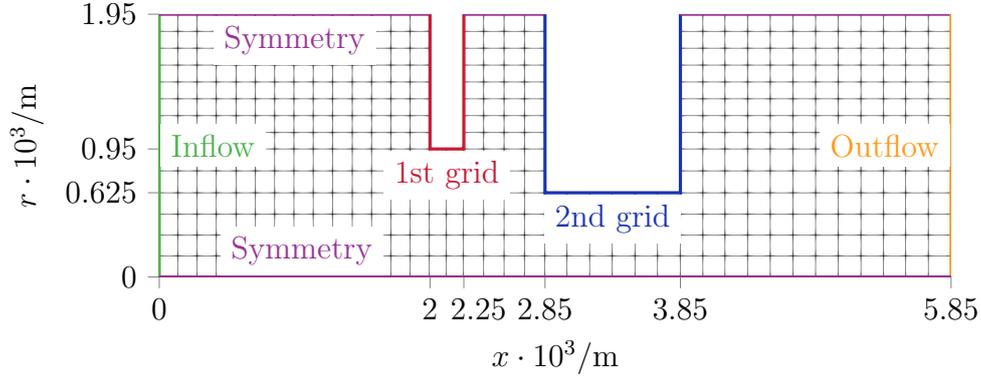
\begin{figure}
    \centering
    \begin{tikzpicture}
\begin{axis}[
width=12cm,
scaled ticks=false,
tick label style={/pgf/number format/fixed},
axis equal image,
axis x line*=bottom,  
axis y line*=left,    
xtick align=outside,    
ytick align=outside,
view={0}{0},
ymin=0,
ymax=468,
xmin=0,
xmax=1400,
xtick={0,
  \pgfkeysvalueof{/pgfplots/xmax}*2/5.85,
  \pgfkeysvalueof{/pgfplots/xmax}*2.25/5.85,
  \pgfkeysvalueof{/pgfplots/xmax}*2.85/5.85,
  \pgfkeysvalueof{/pgfplots/xmax}*3.85/5.85,
  \pgfkeysvalueof{/pgfplots/xmax}
},
xticklabels={0,2,\hspace{11pt}{2.25},2.85,3.85,5.85},
ytick={0,
  \pgfkeysvalueof{/pgfplots/ymax}*0.625/1.95,
  \pgfkeysvalueof{/pgfplots/ymax}*0.95/1.95,
  \pgfkeysvalueof{/pgfplots/ymax}
},
yticklabels={0,0.625,0.95,1.95},
xlabel={$x \cdot 10^3 / \si{\meter}$},
ylabel={$r \cdot 10^3 / \si{\meter}$},
]
\pgfmathsetmacro{\Xa}{2.00/5.85*\pgfkeysvalueof{/pgfplots/xmax}}
\pgfmathsetmacro{\Xb}{2.25/5.85*\pgfkeysvalueof{/pgfplots/xmax}}
\pgfmathsetmacro{\Xc}{2.85/5.85*\pgfkeysvalueof{/pgfplots/xmax}}
\pgfmathsetmacro{\Xd}{3.85/5.85*\pgfkeysvalueof{/pgfplots/xmax}}
\pgfmathsetmacro{\Xe}{\pgfkeysvalueof{/pgfplots/xmax}}
\pgfmathsetmacro{\Ya}{(0.950/1.95)*\pgfkeysvalueof{/pgfplots/ymax}}
\pgfmathsetmacro{\Yb}{(0.625/1.95)*\pgfkeysvalueof{/pgfplots/ymax}}
\pgfmathsetmacro{\Yc}{\pgfkeysvalueof{/pgfplots/ymax}}
\addplot graphics [xmin=0,xmax=\Xe,ymin=0,ymax=\pgfkeysvalueof{/pgfplots/ymax}]{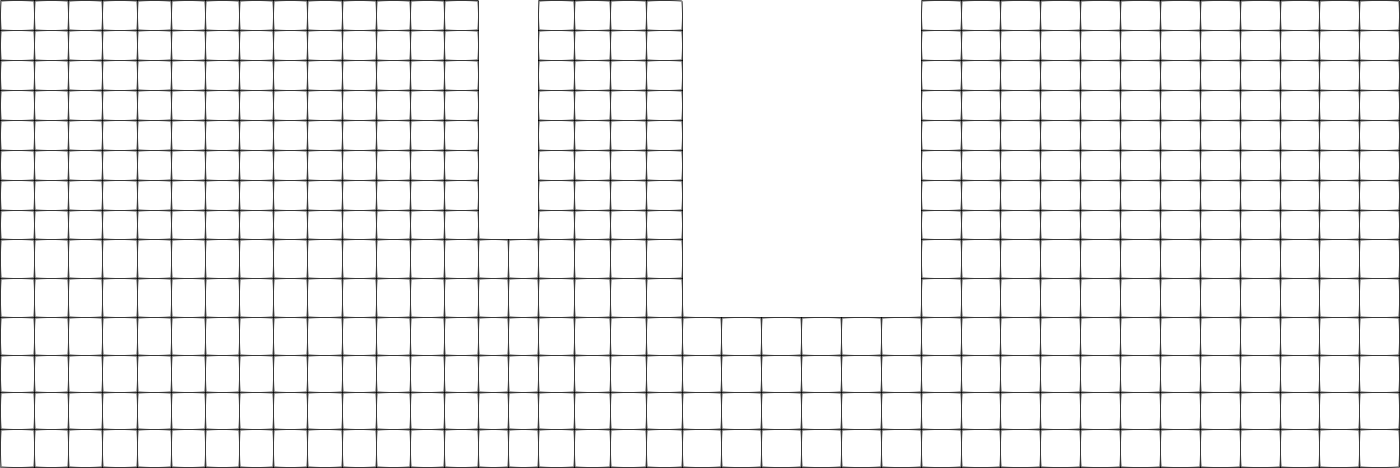};
\draw[col3,very thick] (axis cs:0,\Yc) -- (axis cs:0,0) node[midway,right,fill=white] {Inflow};
\draw[col4,very thick] (0,\Yc) -- (\Xa,\Yc) node[midway,below,fill=white] {Symmetry};
\draw[col2,very thick] (\Xa,\Yc) -- (\Xa,\Ya) -- (\Xb,\Ya) node[midway,below,fill=white] {1st grid} -- (\Xb,\Yc);
\draw[col4,very thick] (\Xb,\Yc) -- (\Xc,\Yc);
\draw[col1,very thick] (\Xc,\Yc) -- (\Xc,\Yb) -- (\Xd,\Yb) node[midway,below,fill=white] {2nd grid}-- (\Xd,\Yc);
\draw[col4,very thick] (\Xd,\Yc) -- (\Xe,\Yc);
\draw[col5,very thick] (\Xe,\Yc) -- (\Xe,0) node[midway,left,fill=white] {Outflow};
\draw[col4,very thick] (\Xe,0) -- (0,0) node[pos=0.82,above,fill=white] {Symmetry};
\end{axis}

\end{tikzpicture}
    \caption{Discretized domain and boundaries used in the simulation. The values of the boundary conditions are summarized in Table \ref{tab:ion-optic-boundaries}.}
    \label{fig:ion-optics-mesh}
\end{figure}

The computational domain and its boundary conditions are illustrated in Figure \ref{fig:ion-optics-mesh} with the corresponding values summarized in Table \ref{tab:ion-optic-boundaries}.
The inflow consists of a quasi-neutral Xenon plasma, with electrons and $\text{Xe}^+$ ions, as detailed in Table \ref{tab:ion-optic-species}.
Given the specified plasma conditions, the resulting electron plasma frequency is approximately $\omega_{\text{e}}\approx\SI{1.78E10}{\per\second}$.
Thus, the simulation time step was set to $\Delta t=\SI{1E-11}{\second}\approx 0.2/\omega_{\text{e}}$ to explicitly resolve the electron plasma oscillations and ensure numerical stability.
A particle weighting of $\omega_p=10^4$ was utilized for both species, resulting in roughly $\SI{4E5}{}$ macro-particles that are predominantly concentrated in the upstream discharge region.
Time-averaging was performed over $N_{\text{avg}}=10^5$ time steps to obtain smooth and statistically converged results for the density distributions.

To determine the spatial distribution of the polynomial degree across the mesh, the iterative modal analysis strategy described in the preceding section was utilized.
Initially, the entire domain was set to a low polynomial degree of $N=1$.
The local solution within each element was then expanded into a hierarchy of orthogonal Legendre polynomials.
If the normalized contribution of the highest active mode exceeded a specified tolerance threshold, the polynomial degree in that element was locally incremented for the subsequent run, up to a maximum limit of $N_{\max}=5$.

This modal refinement indicator was evaluated independently in the axial ($x$) and radial ($r$) directions.
Three distinct physical variables were used for the refinement process: the electric potential $\varphi$, the ion charge density $\rho_i$, and the electron charge density $\rho_e$.
For the smoothly varying electric potential, a constant threshold of $\varepsilon_\varphi=10$ was applied.
However, because the charge densities are evaluated directly from discrete simulation particles, their modal coefficients are subject to statistical noise.
To prevent spurious p-refinement triggered merely by particle noise, the density thresholds were set to $\varepsilon_{\rho_i,k}=200^2\sigma_k^2$ and $\varepsilon_{\rho_e,k}=60^2\sigma_k^2$ for the ions and electrons, respectively.
$\sigma_k^2$ is an estimate of the variance of the normalized contribution of mode $k$ due to statistical fluctuations.
The derivation of this variance is provided in Appendix \ref{app:threshold}.
In theory, a threshold of $(3\sigma_k)^2$ should be sufficient to only trigger refinement within the physical signal.
However, the variance estimate is based on the assumption of independent and identically distributed particle samples, which is not valid for the particle samples in the simulation.
In regions with low particle velocities, single particles can remain in the same element for extended periods, leading to correlated samples and an underestimation of the variance.
This also explains the choice of a higher threshold for the slow ions compared to the faster electrons.

\begin{figure}
    \centering
    \begin{tikzpicture}
\begin{groupplot}[
    group style={
        group size=1 by 4,
        vertical sep=1.0cm,
    },
    width=12cm,
    axis equal image,
    hide axis,
    view={0}{0},
    ymin=0,
    ymax=468,
    xmin=0,
    xmax=1400,
]

\nextgroupplot[title={$N_{\max}=2$}]
\addplot graphics [xmin=0,xmax=1400,ymin=0,ymax=468]{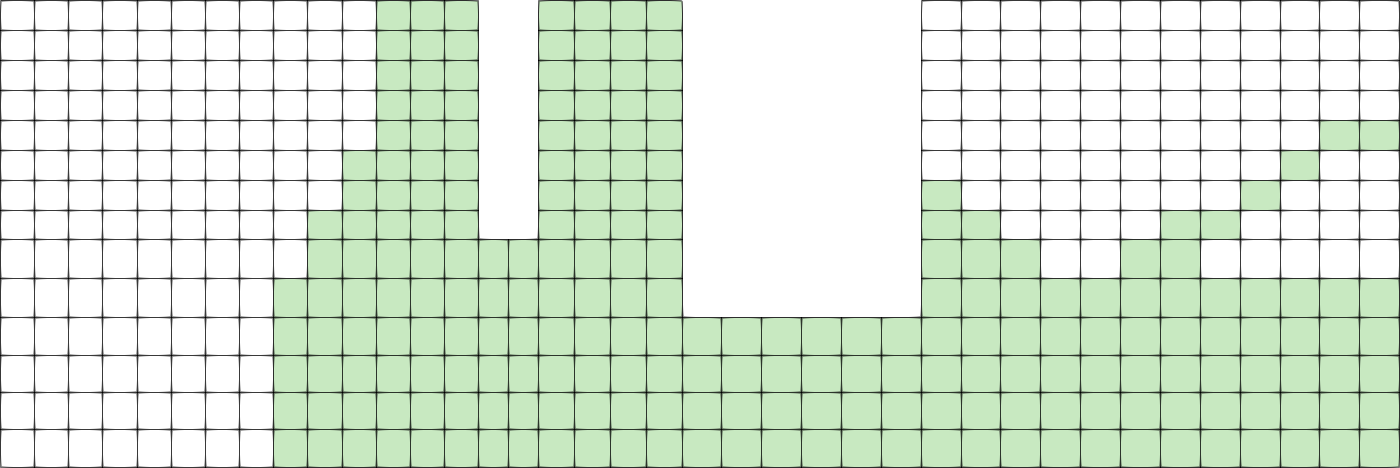};

\nextgroupplot[title={$N_{\max}=3$}]
\addplot graphics [xmin=0,xmax=1400,ymin=0,ymax=468]{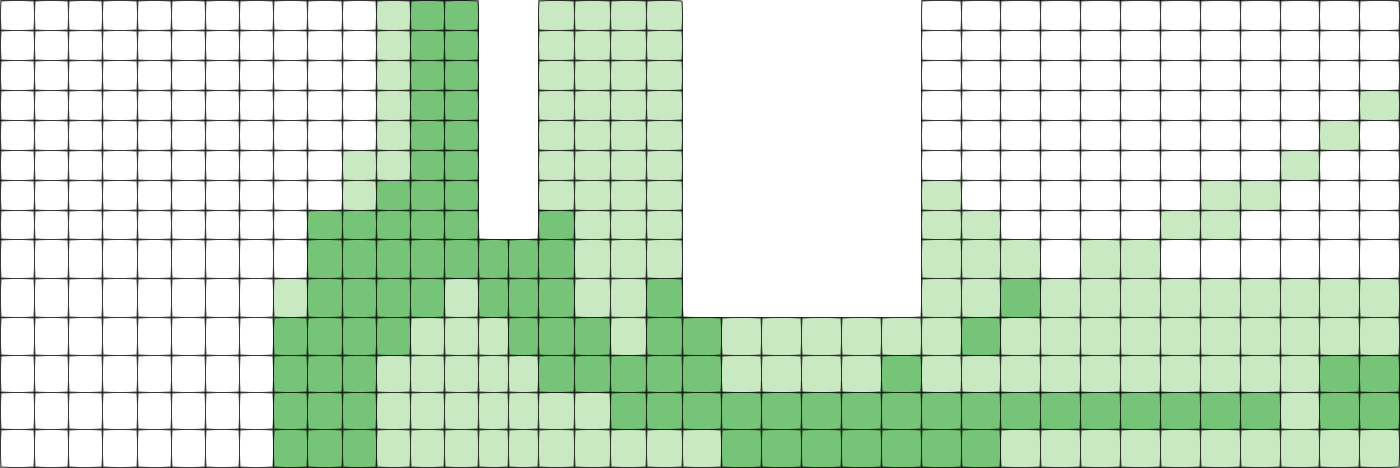};

\nextgroupplot[title={$N_{\max}=4$}]
\addplot graphics [xmin=0,xmax=1400,ymin=0,ymax=468]{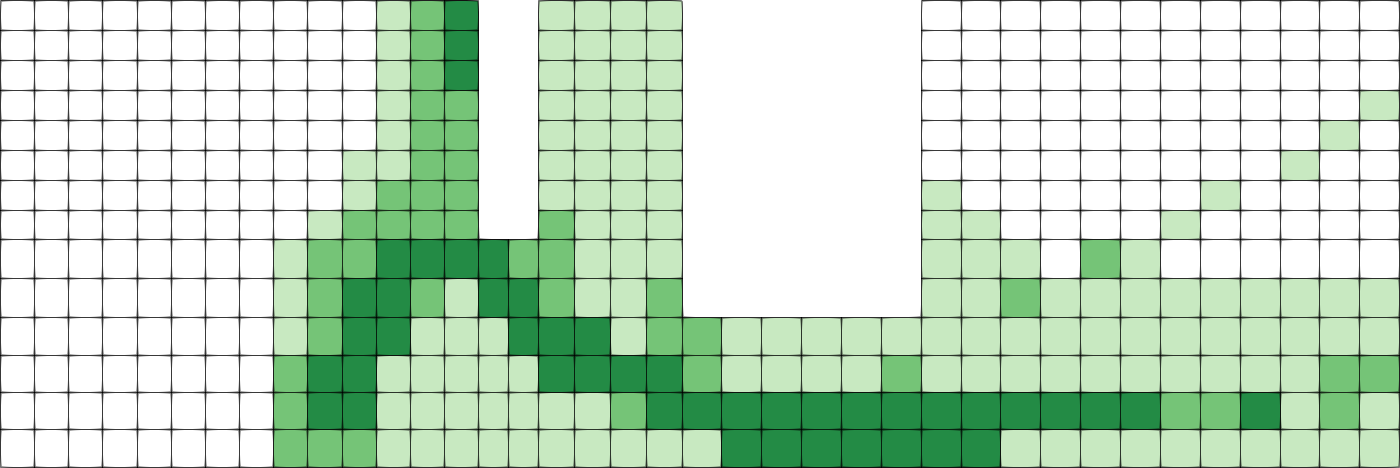};

\nextgroupplot[title={$N_{\max}=5$}]
\addplot graphics [xmin=0,xmax=1400,ymin=0,ymax=468]{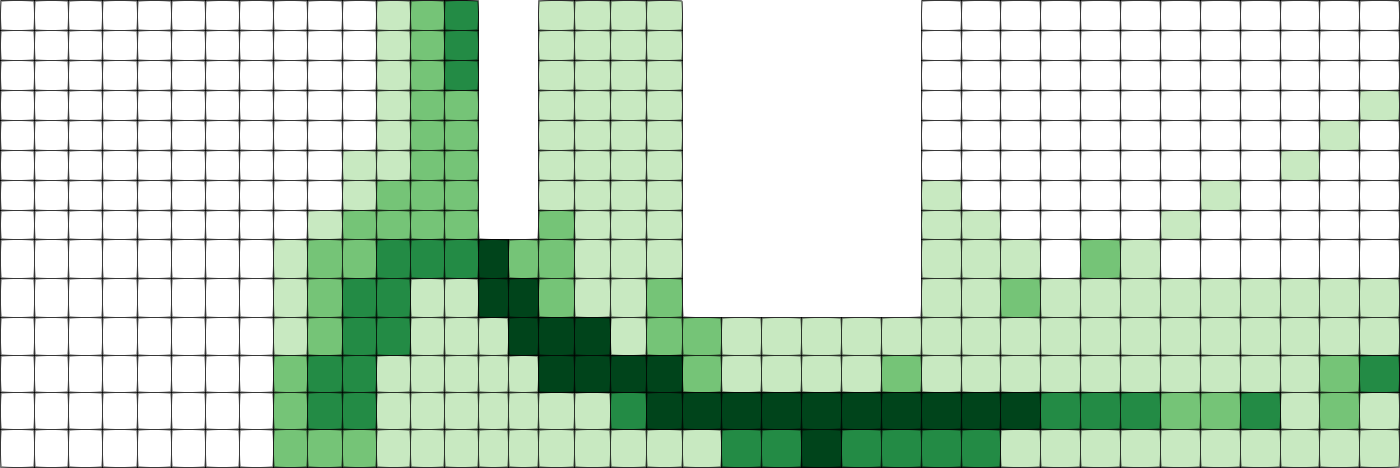};

\end{groupplot}

\path let \p1=(group c1r1.north), \p2=(group c1r4.south) in
    \pgfextra{\pgfmathsetlengthmacro{\colorbarheight}{\y1-\y2}
              \global\let\colorbarheight\colorbarheight};

\begin{axis}[
    at={($(group c1r1.north east)!0.5!(group c1r4.south east)$)},
    anchor=west,
    xshift=0.5cm,
    hide axis,
    scale only axis,
    width=0pt,
    height=0.3*\colorbarheight,
    xmin=0, xmax=1,
    ymin=0, ymax=1,
    colormap={cb}{
        rgb255(0cm)=(255,255,255);rgb255(0.9999cm)=(255,255,255);
        rgb255(1cm)=(200,233,193);rgb255(1.9999cm)=(200,233,193);
        rgb255(2cm)=(117,196,119);rgb255(2.9999cm)=(117,196,119);
        rgb255(3cm)=(35,139,69);rgb255(3.9999cm)=(35,139,69);
        rgb255(4cm)=(0,68,27);rgb255(5cm)=(0,68,27)
    },
    point meta min=0.5,
    point meta max=5.5,
    colorbar right,
    every colorbar/.append style={
        font=\small,
        title={$N$},
        title style={xshift=0pt,yshift=-5pt},
        ytick={1,2,3,4,5},
        yticklabels={1,2,3,4,5},
        ytick style={draw=none},
    },
]
\addplot [draw=none] coordinates {(0,0)};
\end{axis}
\end{tikzpicture}
    \caption{Polynomial degree distribution across the mesh after each successive simulation run. The polynomial degree is increased in the region between the two grids, where strong field and density gradients are present.}
    \label{fig:ion-optic-degree}
\end{figure}

Figure \ref{fig:ion-optic-degree} shows the distribution of the polynomial degree across the mesh after each successive simulation.
After each run, the local solution is analyzed and the polynomial degree is updated.
Between the successive runs, the polynomial degree in each element may either be incremented by one, remain unchanged, or be set to any lower degree, depending on the local solution.

\begin{figure}
    \centering
    \begin{tikzpicture}
\begin{axis}[
width=12cm,
scaled ticks=false,
tick label style={/pgf/number format/fixed},
axis equal image,
axis x line*=bottom,  
axis y line*=left,    
xtick align=outside,    
ytick align=outside,
axis line style={draw=none},
view={0}{0},
ymin=0,
ymax=936,
xmin=0,
xmax=1400,
xtick={0,
  \pgfkeysvalueof{/pgfplots/xmax}*2/5.85,
  \pgfkeysvalueof{/pgfplots/xmax}*2.25/5.85,
  \pgfkeysvalueof{/pgfplots/xmax}*2.85/5.85,
  \pgfkeysvalueof{/pgfplots/xmax}*3.85/5.85,
  \pgfkeysvalueof{/pgfplots/xmax}
},
xticklabels={0,2,\hspace{11pt}{2.25},2.85,3.85,5.85},
ytick={0,
  \pgfkeysvalueof{/pgfplots/ymax}*0.5*(1-0.95/1.95),
  \pgfkeysvalueof{/pgfplots/ymax}*0.5*(1-0.625/1.95),
  \pgfkeysvalueof{/pgfplots/ymax}*0.5,
  \pgfkeysvalueof{/pgfplots/ymax}*0.5*(1+0.625/1.95),
  \pgfkeysvalueof{/pgfplots/ymax}*0.5*(1+0.95/1.95),
  \pgfkeysvalueof{/pgfplots/ymax}
},
yticklabels={-1.95,-0.95,-0.625,0,0.625,0.95,1.95},
xlabel={$x \cdot 10^3 / \si{\meter}$},
ylabel={$r \cdot 10^3 / \si{\meter}$},
colormap={rgb}{rgb255=(0,0,255) rgb255=(0,255,255) rgb255=(0,255,0) rgb255=(255,255,0) rgb255=(255,0,0)},
colorbar,
colorbar style={
	font=\small,
    tick style={color=black},
	anchor=south west,
	at={(rel axis cs:1.05, 0.55)},
	title={$\varphi / \si{\volt}$},
	title style={xshift=0pt,yshift=-5pt},
    height=0.45*\pgfkeysvalueof{/pgfplots/parent axis height},
},
point meta min=-300,
point meta max=1300,
colorbar style={ytick={-300,100,500,900,1300},scaled ticks=false},
]
\pgfmathsetmacro{\Yc}{0.5*\pgfkeysvalueof{/pgfplots/ymax}}
\addplot graphics [xmin=0,xmax=\pgfkeysvalueof{/pgfplots/xmax},ymin=\Yc,ymax=\pgfkeysvalueof{/pgfplots/ymax}]{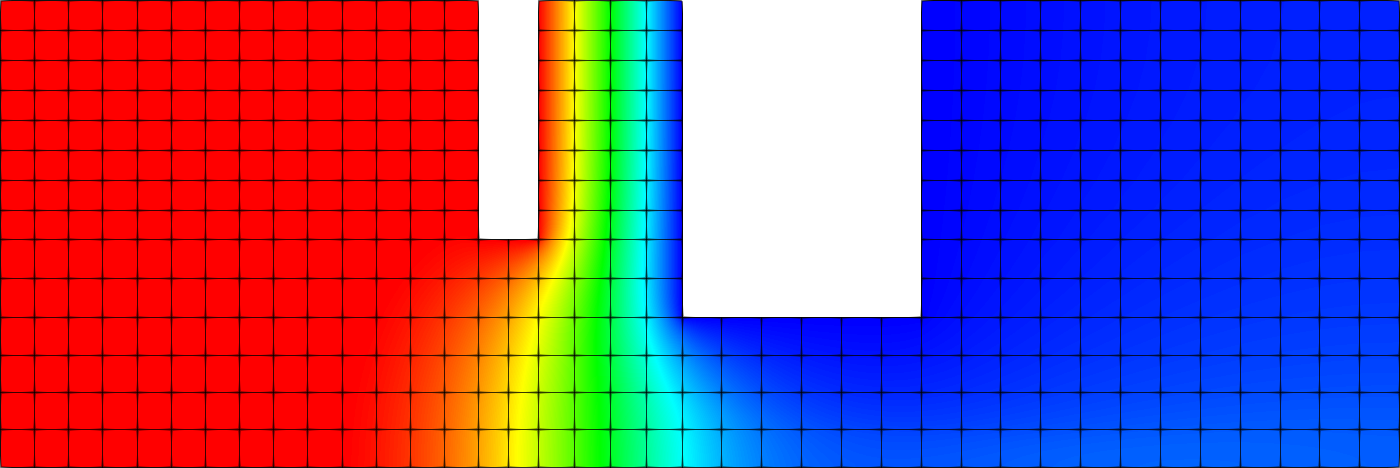};
\end{axis}

\begin{axis}[
width=12cm,
scaled ticks=false,
axis equal image,
view={0}{0},
ymin=0,
ymax=1022,
xmin=0,
xmax=1526,
hide axis,
colormap={rgb}{rgb255=(255,255,255) rgb255=(0,0,255) rgb255=(0,255,255) rgb255=(0,255,0) rgb255=(255,255,0) rgb255=(255,0,0) rgb255=(224,0,255)},
colorbar,
colorbar style={
	font=\small,
    tick style={color=black},
	anchor=north west,
	at={(rel axis cs:1.05, 0.4)},
	title={$\rho_{\text{Xe}^+} / (\si{\coulomb\per\meter\cubed})$},
	title style={xshift=5pt,yshift=-5pt},
    height=0.45*\pgfkeysvalueof{/pgfplots/parent axis height},
},
colorbar style={ytick={0, 0.02, 0.04, 0.06},scaled ticks=false},
point meta min=0,
point meta max=0.06,
]
\pgfmathsetmacro{\Yc}{0.5*\pgfkeysvalueof{/pgfplots/ymax}}
\addplot graphics [xmin=0,xmax=\pgfkeysvalueof{/pgfplots/xmax},ymin=0,ymax=\Yc]{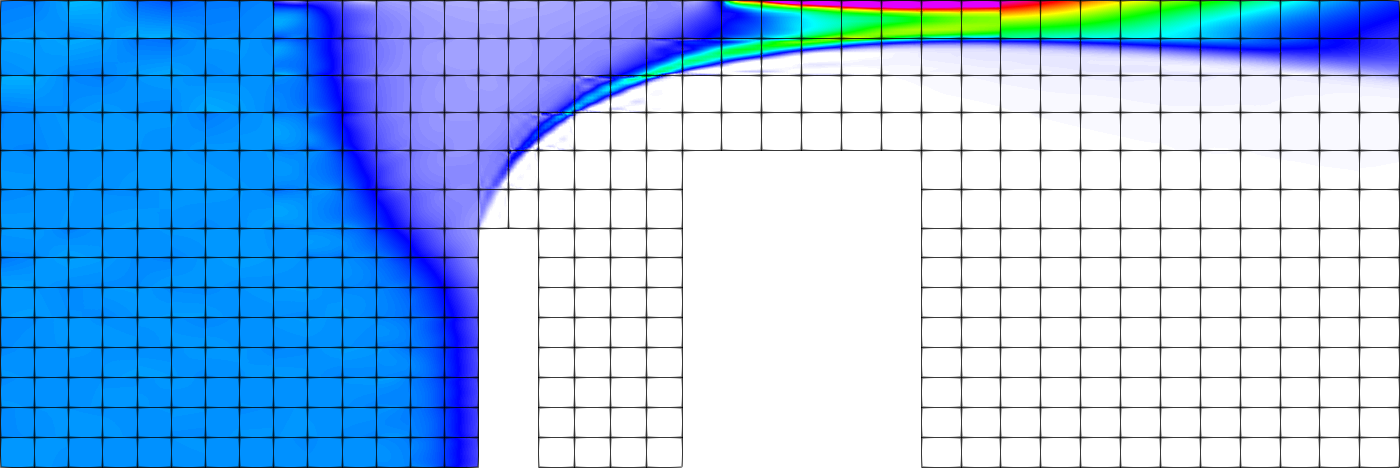};
\end{axis}

\end{tikzpicture}
    \caption{Electric potential (upper plot) and $\text{Xe}^+$ charge density (lower plot) as simulated with the p-adapted HDG-SEM.}
    \label{fig:ion-optics-result}
\end{figure}

The results of the simulation are illustrated in Figure \ref{fig:ion-optics-result}, which shows the electric potential and the $\text{Xe}^+$ ion density distribution within the ion optic.
As illustrated in Figure \ref{fig:ion-optic-degree}, the p-adaptive algorithm effectively identifies the regions of the computational domain that require higher resolution.
In the upstream and bulk plasma region, quasi-neutrality enforces a relatively constant potential, allowing the solver to maintain a low polynomial degree of $N=1$.
However, in the region between the two grids, where the applied grid potentials shield electrons from the plasma and accelerate the ions, strong field and density gradients are present.
In this region, the adaptive solver successfully identifies the unresolved features and locally increases the polynomial degree up to the maximum of $N=5$.
This enables an accurate representation of the solution without the need to uniformly increase the polynomial degree across the entire domain, which would lead to a higher computational cost.
However, in the downstream region in the elements adjacent to the symmetry axis, the polynomial degree remains low despite the presence of strong gradients in the ion density.
This is because the low particle count close to the symmetry axis leads to a high variance estimate, which prevents refinement in this region.
Since the velocities of the ions in this region are relatively high, a lower threshold could be used to trigger refinement.
Since a uniform threshold is employed across the entire domain, however, this would lead to excessive refinement in the upstream region where the ion velocities are low.

\begin{table}
  \caption{Boundary conditions applied in the simulation.}\label{tab:ion-optic-boundaries}
  \begin{center}
    \begin{tabular}{lrr}
      \hline\rule{0pt}{12pt}
      & Field Solver & Particle\\
      \hline\rule{-4pt}{12pt}
      Inflow & $\varphi_\text{In}=\SI{1315}{\volt}$ & inflow, open\\
      1st grid & $\varphi_\text{An}=\SI{1300}{\volt}$ & open\\
      2nd grid & $\varphi_\text{Cat}=\SI{-500}{\volt}$ & open\\
      Outflow & $E_\text{Out}=\SI{0}{\volt\per\meter}$ & open\\
      Symmetry & $E_\text{Sym}=\SI{0}{\volt\per\meter}$ & reflective \\[2pt]
      \hline
    \end{tabular}
  \end{center}
\end{table}

\begin{table}
  \caption{Species-specific parameters and inflow conditions used for the ion optic simulation.}\label{tab:ion-optic-species}
  \begin{center}
    \begin{tabular}{llrr}
      \hline\rule{0pt}{12pt}
      & & Electrons & $\text{Xe}^+$ Ions\\
      \hline\rule{-4pt}{12pt}
      Mass & $m \,/\, \SI{}{\kg}$ & $\SI{9.109E-31}{}$ & $\SI{2.180E-25}{}$\\
      Charge & $q \,/\, \SI{}{\coulomb}$ & $\SI{-1.602E-19}{}$ & $\SI{1.602E-19}{}$\\
      Number density & $n \,/\, \SI{}{\per\meter\cubed}$ & $\SI{1E17}{}$ & $\SI{1E17}{}$\\
      Temperature & $T \,/\, \SI{}{\kelvin}$ & $\SI{40621}{}$ & $\SI{450}{}$\\
      Velocity & $v \,/\, \SI{}{\meter\per\second}$ & $\SI{2500}{}$ & $\SI{2500}{}$\\[2pt]
      \hline
    \end{tabular}
  \end{center}
\end{table}

\section{Conclusion}
This paper presented a p-adaptive high-order hybridizable discontinuous Galerkin spectral element method (HDG-SEM) for solving the Poisson equation within electrostatic Particle-in-Cell plasma simulations.
The method was successfully integrated into the open-source \textsc{PICLas} framework.

Validation against both purely analytical benchmarks (a dielectric sphere) and coupled kinetic phenomena (a one-dimensional plasma sheath) demonstrated that the proposed method maintains the error convergence characteristic of high-order DG schemes.
More importantly, the application of p-adaptation significantly reduced the total number of global degrees of freedom compared to uniformly refined high-order methods resulting in significant memory savings and improved computational efficiency.
Finally, the simulation of an axisymmetric Xenon ion optic verified the robustness of the solver in handling complex, multidimensional geometries with highly localized field gradients.

Future work will focus on a fully dynamic adaptation strategy capable of adjusting polynomial degrees automatically during transient simulations.
In order to improve the approximation of the threshold for the density-based indicator, the variance estimation will be further refined, potentially by assuming non-uniform underlying distributions or by analyzing mode contributions and their variance in a multidimensional manner.
Additionally, the estimation of the number of particle samples within each element might be improved by accounting for the velocity of the particles and their residence time.
Finally, the p-adaptive HDG-SEM solver will be applied to large-scale, 3D unstructured simulations featuring complex chemical reactions and ionization processes.

\section*{Acknowledgments}
This project has received funding from the European Research Council (ERC) under the European Union's Horizon 2020 research and innovation programme (grant agreement No. 899981 MEDUSA).

\appendix

\section{Variance approximation for the modal coefficients} \label{app:threshold}
To construct a threshold for the density-based p-adaptation indicator, we estimate the theoretical variance of the modal coefficients caused by the statistical noise of the particle sampling.
Since the modal analysis is performed separately for each spatial direction, the probability distribution is marginalized into 1D profiles along the respective direction.

Let the marginal probability density function of a physical quantity along a spatial direction (e.g. $x$ or $r$) within an element be given by $f(x)$, where $x\in[x_{\min},x_{\max}]$.
This function can be expanded into a modal basis using orthogonal Legendre polynomials $\psi_k(\xi)$ mapped to the reference element $\xi\in[-1,1]$ as
\begin{equation}
    f(x) = \sum_{k=0}^{N} \hat{f}_k \psi_k(\xi(x)) = \hat{\vc{f}}\cdot\vc{\psi}(\xi(x)).
\end{equation}

To isolate the contribution of each mode, the coefficients $\hat{\vc{f}}$ are obtained by projecting the 1D function onto the modal basis:
\begin{equation}
    \mt{M}\vc{\hat{f}} = \int f(x)\vc{\psi}(\xi(x))\,\text{d}V, \qquad \mt{M} = \int \vc{\psi}(\xi(x))\vc{\psi}(\xi(x))^T\,\text{d}V.
\end{equation}
Transforming the integrals from the physical element to the reference element yields
\begin{equation}
    \int f(x)\vc{\psi}(\xi(x))\,\text{d}V = \int_{-1}^{1} f(x(\xi))\vc{\psi}(\xi)J_x\,\text{d}\xi,
\end{equation}
where $J_x=\int_{-1}^1\int_{-1}^1 J\,\text{d}\eta\,\text{d}\zeta$ is the effective 1D Jacobian and is constant for linear elements.
In axisymmetric simulations, when the spatial direction of interest is the radial direction, the integration must account for the cylindrical volume element.
Consequently, the Jacobian includes the radial distance such that $J_x \propto r$.

Let $X$ be a random variable sampled from the normalized PDF $\bar{f}(\xi) = J_x\,f(x(\xi))$ within the reference element.
In the PIC scheme, this represents the distribution of simulation particles along the respective spatial direction within the element.
The expected value $\vc{\mu}$ and covariance matrix $\vc{\Sigma}$ of the modal contributions are then analytically given by
\begin{align}
    \vc{\mu} &= \mathbb{E}(\vc{\hat{f}}) = \mathbb{E}\left(\mt{M}^{-1}\vc{\psi}(X)\right) = \mt{M}^{-1}\int_{-1}^{1} \vc{\psi}(\xi)\bar{f}(\xi)\,\text{d}\xi, \\
    \mt{\Sigma} &= \text{Cov}(\vc{\hat{f}}) = \text{Cov}\left(\mt{M}^{-1}\vc{\psi}(X)\right) = \mt{M}^{-1}\left(\int_{-1}^{1} \vc{\psi}(\xi)\vc{\psi}(\xi)^T\bar{f}(\xi)\,\text{d}\xi\right)\mt{M}^{-T} - \vc{\mu}\vc{\mu}^T.
\end{align}

To obtain a practical estimate of the noise variance, we assume that the underlying physical distribution is approximately uniform across the element with volume $|\Omega|$, $f(x) = 1/|\Omega|$.
Normalizing by the element volume, the variance of the $k$-th modal coefficient induced by statistical noise from $N_p$ independent particles is:
\begin{equation}
    \sigma_k^2 = \frac{|\Omega|^2}{N_p}\Sigma_{kk}.
\end{equation}
In the implementation, $N_p$ is estimated using the total number of discrete computational particle samples within the element during the sampling period.

\bibliographystyle{plain}
\bibliography{references}

\end{document}